\documentclass[12pt]{article}
\usepackage[margin=1in]{geometry}
\usepackage{amsmath}
\usepackage{setspace}
\usepackage{graphicx}
\usepackage{epstopdf}
\usepackage{hyperref}

\title{Robustness of ANCOVA in randomised trials with unequal randomisation}

\newcommand{\etal}{\textit{et al }}

\newcommand\ci{\perp\!\!\!\perp}

\newtheorem{thm}{Theorem}

\author{JONATHAN W. BARTLETT \\
Department of Mathematical Sciences, University of Bath, Bath, BA2 7AY, UK \\
email: j.w.bartlett@bath.ac.uk \\
ORCID ID: 0000-0001-7117-0195}

\begin{document}
\maketitle

\large

\abstract{Randomised trials with continuous outcomes are often analysed using ANCOVA, with adjustment for prognostic baseline covariates. In an article published recently, Wang \etal proved that in this setting the model based standard error estimator for the treamtent effect is consistent under outcome model misspecification, provided the probability of randomisation to each treatment is 1/2. In this article, we extend their results allowing for unequal randomisation. These demonstrate that the model based standard error is in general inconsistent when the randomisation probability differs from 1/2. In contrast, the sandwich standard error can provide asymptotically valid inferences under misspecification when randomisation probabilities are not equal, and is therefore recommended when randomisation is unequal.}


\section{Introduction}
In randomised trials with continuous outcomes the baseline covariate adjusted treatment effect estimator is consistent even if the assumed linear regression model (ANCOVA) is misspecified \cite{Yang2001}. Recently Wang \etal proved that under certain conditions, the model based variance estimator from an ANCOVA analysis of a randomised trial is valid under arbitrary misspecification, and therefore advocated its use for analysis of trials with continuous outcomes \cite{wang2019analysis}. Concurrently, the US FDA have recently issued draft guidance on the topic of baseline covariate adjustment in randomised trials with continuous outcomes \cite{fda2019cts}. This draft guidance also advocates use of ANCOVA, and states that the type I error rate is controlled even when the model is misspecified.

An assumption used by Wang \etal is that the probabilities of randomisation to the two arms are equal \cite{wang2019analysis}. While this is commonly the case in randomised trials, many trials are conducted with unequal randomisation probabilities. In particular often the probability of randomisation to the experimental arm is greater than 1/2 in light of a hoped for improved outcome on the experimental treatment compared to control. In this article we explore the impact of violations of the equal randomisation probability assumption on the validity of the model based ANCOVA standard error, and thereby the impact on type I error and confidence interval coverage.

\section{Model based ANCOVA variance estimation with unequal randomisation}
Following the notation of Wang \etal, we assume we observe $n$ i.i.d. copies of $(\mathbf W, A, Y)$, where $\mathbf W$ is a $k\times 1$ column vector of bounded baseline covariates, $A$ is the binary treatment group indicator ($A=1$ for experimental treatment, $A=0$ for control) and $Y$ is the continuous outcome. Like Wang \etal, we assume $A \ci W$, but we let $P(A=1)=\pi$, where $\pi$ may differ from 1/2.

The target of inference is the average treatment effect $\Delta=E(Y|A=1)-E(Y|A=0)$. The unadjusted estimator of $\Delta$ is the difference in treatment group sample means: $\hat{\Delta}^{unadj} = \sum^{n}_{i=1} Y_{i}A_{i} / \sum^{n}_{i=1} A_{i} - \sum^{n}_{i=1} Y_{i}(1-A_{i}) / \sum^{n}_{i=1} (1-A_{i})$. The ANCOVA estimator adjusts for the baseline covariates $\mathbf W$ by fitting the following linear regression model:
\begin{eqnarray}
E(Y|A,\mathbf W) = \beta_{0} + \beta_{A} A + \beta^{T}_{\mathbf W} \mathbf W
\label{ancovaLinearModel}
\end{eqnarray}
where the regression coefficients are estimated by the ordinary least square estimators $\hat{\beta}_{0}$, $\hat{\beta}_{A}$, and $\hat{\beta}_{\mathbf W}$. The ANCOVA estimator $\hat{\Delta}^{ancova}$ of $\Delta$ is $\hat{\Delta}^{ancova}=\hat{\beta}_{A}$. We let $\underline{\beta}_{0}$, $\underline{\beta}_{A}$ and $\underline{\beta}_{W}$ denote the probability limits of these estimators.

As noted by Wang \etal, Yang \& Tsiatis \cite{Yang2001} and Tsiatis \etal \cite{Tsiatis2008} proved, under the stated assumptions, that $\hat{\Delta}^{ancova}$ is a consistent estimator of $\Delta$ under arbitrary misspecification of the linear model in equation \eqref{ancovaLinearModel}, so that $\underline{\beta}_{A}=\Delta$. Following Wang \etal, we let $Var^{*}(\hat{\Delta}^{ancova})$ denote the asymptotic variance of $\hat{\Delta}^{ancova}$, in the sense that $n^{1/2}(\hat{\Delta}^{ancova}-\Delta)$ converges in distribution to a mean zero normal with variance $Var^{*}(\hat{\Delta}^{ancova})$.

Inferences from ANCOVA are by default in statistical software packages based on the so called model based variance estimator for $\hat{\Delta}^{ancova}$, which is given by
\begin{eqnarray*}
\widehat{Var}(\hat{\Delta}^{ancova}) = \frac{\widehat{Var}(Y-\hat{\beta}_{0}-\hat{\beta}_{A} A - \hat{\beta}^{T}_{\mathbf W} \mathbf W)}{(n-1) \left[\widehat{Var}(A) - \widehat{Cov}(\mathbf W,A)^{T} \widehat{Var}(\mathbf W)^{-1} \widehat{Cov}(\mathbf W, A) \right]}
\end{eqnarray*}
where following Wang \etal the estimated variances and covariances on the right hand side are sample variance and sample covariances, with degrees of freedom taken into account (see the Supporting Information of Wang \etal \cite{wang2019analysis} for precise definitions). Wang \etal prove that when $\pi=1/2$, $n \widehat{Var}(\hat{\Delta}^{ancova})$ converges in probability to the true asymptotic variance $Var^{*}(\hat{\Delta}^{ancova})$. As a consequence, under these assumptions, asymptotically Wald-type hypothesis tests have the correct type I error under the null $\Delta=0$ and the corresponding confidence intervals attain their nominal coverage levels.

The following theorem, proved in the Supporting Information, gives the asymptotic variance of $\hat{\Delta}^{ancova}$ for arbitrary $0<\pi<1$, generalising the results of Wang \etal.

\begin{thm}
Given the previously stated assumptions with $0<\pi<1$, the true asymptotic variance $Var^{*}(\hat{\Delta}^{ancova})$ of the ANCOVA estimator $\hat{\Delta}^{ancova}$ is given by
\begin{eqnarray*}
Var^{*}(\hat{\Delta}^{ancova}) = \frac{Var(Y-\underline{\beta}^{T}_{\mathbf W} \mathbf W | A=1)}{\pi} + \frac{Var(Y-\underline{\beta}^{T}_{\mathbf W} \mathbf W | A=0)}{1-\pi} 
\end{eqnarray*}
\end{thm}

The next theorem, again proved in the Supporting Information,, gives the probability limit of $n \widehat{Var}(\hat{\Delta}^{ancova})$ under arbitrary $0<\pi<1$.
\begin{thm}
For the model based variance estimator $\widehat{Var}(\hat{\Delta}^{ancova})$ we have
\begin{eqnarray*}
n \widehat{Var}(\hat{\Delta}^{ancova}) \xrightarrow{P} \frac{Var(Y-\underline{\beta}^{T}_{\mathbf W} \mathbf W | A=1)}{1-\pi} + \frac{Var(Y-\underline{\beta}^{T}_{\mathbf W} \mathbf W | A=0)}{\pi} 
\end{eqnarray*}
\end{thm}

Together the two theorems imply that the model based variance estimator of $\hat{\Delta}^{ancova}$ is only asymptotically valid (and hence hypothesis tests and confidence intervals have correct asymptotic size and coverage) if $\pi=1/2$, as assumed by Wang \etal, or if $Var(Y-\underline{\beta}^{T}_{\mathbf W} \mathbf W | A=1)=Var(Y-\underline{\beta}^{T}_{\mathbf W} \mathbf W | A=0)$. When $\pi \neq 1/2$, the latter conditional variances are not in general equal under misspecification of the outcome model. For example, even if the conditional mean function $E(Y|A,\mathbf W)$ is correctly specified, if the conditional variance of $Y$ given $W$ in the two treatment groups differ, the model based ANCOVA variance estimator is biased. Alternatively, even if $Var(Y|A=1)=Var(Y|A=0)$, if $Cov(Y,\underline{\beta}^{T}_{\mathbf W} \mathbf W |A=1) \neq Cov(Y,\underline{\beta}^{T}_{\mathbf W} \mathbf W |A=0)$, the model based ANCOVA variance estimator is again biased. This would in general occur if the outcome had the same variance in the two treatment groups, but the covariates $\mathbf W$ were prognostic for $Y$ to different extents in the two treatment groups.

We note that a special case of our result occurs when $W$ is empty, such that $\hat{\Delta}^{ancova}=\hat{\Delta}^{unadj}$. In this case our result corresponds to the well known fact that the two sample t-test does not control the type I error rate in general if the outcome variable has different variance in the two groups, which leads to Welch's adaptation of the t-test allowing for unequal variances.

Our results imply that when $\pi \neq 1/2$, the model based ANCOVA variance estimator could be biased downwards or upwards, depending on the configuration, leading to a type I error rate either below or above the nominal level. Suppose for example that $\pi>1/2$, such that a greater proportion of patients are randomised to the experimental treatment. Then if $Var(Y-\underline{\beta}^{T}_{\mathbf W} \mathbf W | A=1)>Var(Y-\underline{\beta}^{T}_{\mathbf W} \mathbf W | A=0)$ the model based ANCOVA variance is too large, leading to type I error rates lower than the nominal level, whereas if $Var(Y-\underline{\beta}^{T}_{\mathbf W} \mathbf W | A=1)<Var(Y-\underline{\beta}^{T}_{\mathbf W} \mathbf W | A=0)$ the model based ANCOVA variance is too small, leading to inflated type I error rates.

\section{Discussion}
We have shown that the model based ANCOVA variance estimator of the average treatment effect is under general misspecification of the outcome model inconsistent when $\pi \neq 0.5$. In trials with unequal randomisation this variance estimator cannot therefore be recommended for general use. Instead, the sandwich variance estimator, as described by Tsiatis \etal \cite{Tsiatis2008}, provides asymptotically valid inferences for any randomisation probability under arbitrary misspecification. An important exception is if randomisation is not simple, as was assumed here and in Wang \etal \cite{wang2019analysis}. For example, as noted by Wang \etal, under stratified randomisation schemes, obtaining asymptotically valid standard errors when covariates not used in the randomisation are adjusted for, under general misspecification of the outcome model, remains an open problem.

\section*{Acknowledgments}
The author thanks David Wright and Daniel Jackson for useful discussions on the topic.

\section*{Supporting Information}
We prove Theorems 1 and 2 of the main paper, referring frequently to the supporting information of Wang \etal \cite{wang2019analysis}.

\subsection*{Proof of Theorem 1}
Following the proof of Theorem B.2 of Wang \etal, the estimating function corresponding to the ANCOVA regression is given by
\begin{eqnarray*}
\psi_{\beta}(Y,A,\mathbf W) = ( Y - \beta_{0} - \beta_{A} A - \beta^{T}_{\mathbf W} \mathbf W) \begin{pmatrix} 1 \\ A \\ W \end{pmatrix}
\end{eqnarray*}
Then as noted by Wang \etal, the OLS estimators $\hat{\beta}=(\hat{\beta}_{0},\hat{\beta}_{A},\hat{\beta}^{T}_{\mathbf W})$ are the solutions to the estimating equation $\sum^{n}_{i=1} \psi_{\hat{\beta}}(Y,A,\mathbf W) = 0$ and its probability limit $\underline{\beta}$ satisfies $E(\psi_{\underline{\beta}}(Y,A,\mathbf W)) =0$. The influence function of $\hat{\beta}$ is
\begin{eqnarray*}
IF_{\hat{\beta}}(Y,A,W) = - \left[ E\left(\frac{\partial \psi_{\underline{\beta}}(Y,A,W)}{\partial \underline{\beta}^{T}} \right) \right]^{-1} \psi_{\underline{\beta}}(Y,A,W)
\end{eqnarray*}
After some matrix algebra, and using the fact that $A \ci W$, one can show that the influence function of $\hat{\Delta}^{ancova}$ is
\begin{eqnarray*}
IF_{ancova}(Y,A,W) = \frac{A-\pi}{\pi(1-\pi)} ( Y - \underline{\beta}_{0} - \underline{\beta}_{A} A - \underline{\beta}^{T}_{\mathbf W} \mathbf W)
\end{eqnarray*}
Note that when $\pi=1/2$, this reduces to the corresponding expression given by Wang \etal. The asymptotic variance of the estimator $\hat{\Delta}^{ancova}$ is then given by the variance of this influence function. Since influence functions have mean zero, this variance is given by
\begin{eqnarray*}
Var(IF_{ancova}(Y,A,W)) &=& E\left[\frac{(A-\pi)^{2}}{\pi^{2}(1-\pi)^{2}} ( Y - \underline{\beta}_{0} - \underline{\beta}_{A} A - \underline{\beta}^{T}_{\mathbf W} \mathbf W)^{2} \right] 
\end{eqnarray*}
Then using the fact that $E(\psi_{\underline{\beta}}(Y,A,\mathbf W)) =0$, we have that $E\left(Y - \underline{\beta}_{0} - \underline{\beta}_{A} A - \underline{\beta}^{T}_{\mathbf W} \mathbf W\right) =0$, and thus that 
\begin{eqnarray*}
Var(IF_{ancova}(Y,A,W)) &=& \frac{1}{\pi^{2}(1-\pi)^{2}} \biggl[ \pi(1-\pi)^{2} Var(Y - \underline{\beta}_{0} - \underline{\beta}_{A} A - \underline{\beta}^{T}_{\mathbf W} \mathbf W | A=1) \\
&& + (1-\pi)\pi^{2} Var(Y - \underline{\beta}_{0} - \underline{\beta}_{A} A - \underline{\beta}^{T}_{\mathbf W} \mathbf W | A=0) \biggl] \\
&=& \frac{1}{\pi^{2}(1-\pi)^{2}} \biggl[ \pi(1-\pi)^{2} Var(Y  - \underline{\beta}^{T}_{\mathbf W} \mathbf W | A=1)  \\
&& + (1-\pi)\pi^{2} Var(Y - \underline{\beta}^{T}_{\mathbf W} \mathbf W | A=0) \biggl] \\
&=& \frac{Var(Y  - \underline{\beta}^{T}_{\mathbf W} \mathbf W | A=1)}{\pi} + \frac{Var(Y - \underline{\beta}^{T}_{\mathbf W} \mathbf W | A=0)}{1-\pi}
\end{eqnarray*}
as required.

\subsection*{Proof of Theorem 2}
Theorem B.3 of Wang \etal argues why $\widehat{Var}(Y - \hat{\beta}_{0} - \hat{\beta}_{A} A - \hat{\beta}^{T}_{\mathbf W} \mathbf W) \xrightarrow{P} Var(Y - \underline{\beta}_{A} A - \underline{\beta}^{T}_{\mathbf W} \mathbf W)$, and their argument applies for any $0<\pi<1$. Next, we have that $n/(n-1) \rightarrow 1$, $\widehat{Var}(A) \xrightarrow{P} Var(A)=\pi(1-\pi)$, by independence of $A$ and $\mathbf W$ $\widehat{Cov}(\mathbf W, A) \xrightarrow{P} Cov(\mathbf W,A) = \mathbf 0$, and $\widehat{Var}(\mathbf W) \xrightarrow{P} Var(\mathbf W)$. Then from the definition of $\widehat{Var}(\hat{\Delta}^{ancova})$ it follows that
\begin{eqnarray*}
n \widehat{Var}(\hat{\Delta}^{ancova}) \xrightarrow{P} \frac{Var(Y - \underline{\beta}_{A} A - \underline{\beta}^{T}_{\mathbf W} \mathbf W)}{\pi(1-\pi)}
\end{eqnarray*}
Next, we write the variance in the numerator as
\begin{eqnarray*}
Var(Y - \underline{\beta}_{A} A - \underline{\beta}^{T}_{\mathbf W} \mathbf W) = Var(E(Y - \underline{\beta}_{A} A - \underline{\beta}^{T}_{\mathbf W} \mathbf W | A)) + E(Var(Y - \underline{\beta}_{A} A - \underline{\beta}^{T}_{\mathbf W} \mathbf W | A))
\end{eqnarray*}
The second of these terms can be expressed as
\begin{eqnarray*}
E(Var(Y - \underline{\beta}_{A} A - \underline{\beta}^{T}_{\mathbf W} \mathbf W | A)) &=& \pi Var(Y - \underline{\beta}_{A} A - \underline{\beta}^{T}_{\mathbf W} \mathbf W | A=1) \\
&& + (1-\pi) Var(Y - \underline{\beta}_{A} A - \underline{\beta}^{T}_{\mathbf W} \mathbf W | A=0) \\
 &=& \pi Var(Y - \underline{\beta}^{T}_{\mathbf W} \mathbf W | A=1) + (1-\pi) Var(Y - \underline{\beta}^{T}_{\mathbf W} \mathbf W | A=0)
\end{eqnarray*}
Next, the fact that $E(\psi_{\underline{\beta}}(Y,A,\mathbf W)) =0$ implies
\begin{eqnarray*}
E\left[ (Y - \underline{\beta}_{0} - \underline{\beta}_{A} A - \underline{\beta}^{T}_{\mathbf W} \mathbf W) A \right] = 0
\end{eqnarray*}
which in turn implies
\begin{eqnarray*}
\pi E\left[ Y - \underline{\beta}_{0} - \underline{\beta}_{A} A - \underline{\beta}^{T}_{\mathbf W} \mathbf W | A=1 \right] = 0
\end{eqnarray*}
Then the fact that $E\left(Y - \underline{\beta}_{0} - \underline{\beta}_{A} A - \underline{\beta}^{T}_{\mathbf W} \mathbf W\right) =0$ means that
\begin{eqnarray*}
\pi E(Y - \underline{\beta}_{0} - \underline{\beta}_{A} A - \underline{\beta}^{T}_{\mathbf W} \mathbf W | A=1) + (1-\pi) E(Y - \underline{\beta}_{0} - \underline{\beta}_{A} A - \underline{\beta}^{T}_{\mathbf W} \mathbf W | A=0) = 0
\end{eqnarray*}
and since $\pi E\left[ Y - \underline{\beta}_{0} - \underline{\beta}_{A} A - \underline{\beta}^{T}_{\mathbf W} \mathbf W | A=1 \right] = 0$, we have that
\begin{eqnarray*}
E\left[ Y - \underline{\beta}_{0} - \underline{\beta}_{A} A - \underline{\beta}^{T}_{\mathbf W} \mathbf W | A=0 \right] = 0
\end{eqnarray*}
Thus we have shown that 
\begin{eqnarray*}
Var(E(Y - \underline{\beta}_{0} - \underline{\beta}_{A} A - \underline{\beta}^{T}_{\mathbf W} \mathbf W | A)) = 0
\end{eqnarray*}
and therefore also that
\begin{eqnarray*}
Var(E(Y - \underline{\beta}_{A} A - \underline{\beta}^{T}_{\mathbf W} \mathbf W | A)) &=& Var(\underline{\beta}_{0} + E(Y - \underline{\beta}_{0} - \underline{\beta}_{A} A - \underline{\beta}^{T}_{\mathbf W} \mathbf W | A)) \\
&=& Var(E(Y - \underline{\beta}_{0} - \underline{\beta}_{A} A - \underline{\beta}^{T}_{\mathbf W} \mathbf W | A)) \\
&=& 0
\end{eqnarray*}
We have thus shown
\begin{eqnarray*}
Var(Y - \underline{\beta}_{A} A - \underline{\beta}^{T}_{\mathbf W} \mathbf W) = \pi Var(Y - \underline{\beta}^{T}_{\mathbf W} \mathbf W | A=1) + (1-\pi) Var(Y - \underline{\beta}^{T}_{\mathbf W} \mathbf W | A=0)
\end{eqnarray*}
and so
\begin{eqnarray*}
n \widehat{Var}(\hat{\Delta}^{ancova}) & \xrightarrow{P} & \frac{\pi Var(Y - \underline{\beta}^{T}_{\mathbf W} \mathbf W | A=1) + (1-\pi) Var(Y - \underline{\beta}^{T}_{\mathbf W} \mathbf W | A=0))}{\pi(1-\pi)} \\
&=& \frac{Var(Y - \underline{\beta}^{T}_{\mathbf W} \mathbf W | A=1)}{1-\pi} + \frac{Var(Y - \underline{\beta}^{T}_{\mathbf W} \mathbf W | A=0)}{\pi}
\end{eqnarray*}
as was required to be shown.


\begin{thebibliography}{1}

\bibitem{fda2019cts}
U.S. Food and Drug Administration.
\newblock {Adjusting for Covariates in Randomized Clinical Trials for Drugs and
  Biologics with Continuous Outcomes}.
\newblock
  \url{https://www.fda.gov/regulatory-information/search-fda-guidance-documents/adjusting-covariates-randomized-clinical-trials-drugs-and-biologics-continuous-outcomes-guidance},
  2019.

\bibitem{Tsiatis2008}
Anastasios~A Tsiatis, Marie Davidian, Min Zhang, and Xiaomin Lu.
\newblock Covariate adjustment for two-sample treatment comparisons in
  randomized clinical trials: a principled yet flexible approach.
\newblock {\em Statistics in Medicine}, 27(23):4658--4677, 2008.

\bibitem{wang2019analysis}
Bingkai Wang, Elizabeth~L Ogburn, and Michael Rosenblum.
\newblock Analysis of covariance (ancova) in randomized trials: More precision
  and valid confidence intervals, without model assumptions.
\newblock {\em Biometrics}, 2019.

\bibitem{Yang2001}
Li~Yang and Anastasios~A Tsiatis.
\newblock Efficiency study of estimators for a treatment effect in a
  pretest--posttest trial.
\newblock {\em The American Statistician}, 55(4):314--321, 2001.

\end{thebibliography}
\end{document}